\documentclass[twocolumn,preprintnumbers,superscriptaddress,prl,nofootinbib,longbibliography,amsmath,amssymb]{revtex4-1}

\usepackage{graphicx}
\usepackage{dcolumn}
\usepackage{bm}
\usepackage{color}
\usepackage{ulem}
\usepackage{gensymb}
\usepackage{braket}
\usepackage{amsmath}
\usepackage[percent]{overpic}

\begin{document}

\title{The sample-dependent and sample-independent thermal transport properties of $\alpha$-RuCl$_3$}

\author{Heda Zhang}
\affiliation{Materials Science and Technology Division, Oak Ridge National Laboratory, Oak Ridge, Tennessee 37831, USA}

\author{Andrew F May}
\affiliation{Materials Science and Technology Division, Oak Ridge National Laboratory, Oak Ridge, Tennessee 37831, USA}

\author{Hu Miao}
\affiliation{Materials Science and Technology Division, Oak Ridge National Laboratory, Oak Ridge, Tennessee 37831, USA}

\author{Brian C Sales}
\affiliation{Materials Science and Technology Division, Oak Ridge National Laboratory, Oak Ridge, Tennessee 37831, USA}

\author{David G Mandrus}
\affiliation{Materials Science and Technology Division, Oak Ridge National Laboratory, Oak Ridge, Tennessee 37831, USA}

\author{Stephen E Nagler}
\affiliation{Neutron Scattering Division, Oak Ridge National Laboratory, Oak Ridge, Tennessee 37831, USA}

\author{Michael A McGuire}
\affiliation{Materials Science and Technology Division, Oak Ridge National Laboratory, Oak Ridge, Tennessee 37831, USA}

\author{Jiaqiang Yan}
\email{yanj@ornl.gov}
\affiliation{Materials Science and Technology Division, Oak Ridge National Laboratory, Oak Ridge, Tennessee 37831, USA}

\date{\today}

\begin{abstract}
We investigated the thermal transport properties of two $\alpha$-RuCl$_3$ crystals with different degrees of stacking disorder to understand the origin of the previously reported oscillatory feature in the field dependence of thermal conductivity. Crystal I shows only one magnetic order around 13\,K, which is near the highest T$_N$ for $\alpha$-RuCl$_3$ with stacking faults. Crystal II has less stacking disorder, with a dominant heat capacity at 7.6\,K along with weak anomalies at 10\,K and 13\,K. In the temperature and field dependence of thermal conductivity, no obvious anomaly was observed to be associated with the magnetic order around 13\,K for either crystal or around 10\,K for crystal II. Crystal II, with less disorder, showed clear oscillations in the field dependence of thermal conductivity, while crystal I, with more disorder, did not. For crystal I, an L-shaped region in the temperature-field space was observed  where thermal Hall conductivity $\kappa_{xy}$/T is within $\pm$20\% of the half quantized thermal Hall conductivity $\kappa_{HQ}$/T. While for crystal II, $\kappa_{xy}$/T  reaches $\kappa_{HQ}$/T  only in the high field and high temperature regime with no indication of a plateau at $\kappa_{HQ}$/T. Our thermal conductivity data suggest the oscillatory features are inherent to the zig-zag ordered phase with T$_N$ near 7\,K. Our planar thermal Hall effect measurements highlight the sensitivity of this phenomena to stacking disorder. Overall, our results highlight the importance of understanding and controlling crystallographic disorder for obtaining and interpreting intrinsic thermal transport properties in $\alpha$-RuCl$_3$.

 \end{abstract}

\maketitle

In the last decade, $\alpha$-RuCl$_3$ was intensively studied as a promising candidate material for realizing Kitaev quantum spin liquids that have Majorana fermions as the elementary excitation\cite{takagi2019concept}. $\alpha$-RuCl$_3$ is a cleavable, layered magnetic material with the van der Waals bonded honeycomb layers formed by edge sharing RuCl$_6$ octahedra\cite{mcguire2017crystal}.  Below T$_N\,\approx$\,7\,K, $\alpha$-RuCl$_3$ shows a zig-zag type magnetic order\cite{plumb2014alpha}. However, this magnetic order can be suppressed by applying an in-plane magnetic field above $\approx$70\,kOe. A field-induced quantum spin liquid state is proposed in the intermediate field range before getting to the field polarized state at even higher fields. Recently, there are two fascinating observations on the thermal transport properties of $\alpha$-RuCl$_3$ in the field-induced quantum spin liquid state. The first one is the observed half-integer quantized thermal Hall conductance which is believed to be one of the fingerprints for Majorana fermions of the fractionalized spin excitations in $\alpha$-RuCl$_3$\cite{kasahara2018majorana}. While some groups reported the plateau like feature at half quantized value in a certain temperature and field range, other groups observed a strongly temperature dependent thermal Hall conductance and proposed a bosonic origin of the observed thermal Hall effect\cite{yokoi2021half,yamashita2020sample,bruin2022robustness, hentrich2019large,lefranccois2022evidence,czajka2022planar}. The other intriguing experimental observation is the oscillatory features of the longitudal thermal conductivity as a function of in-plane magnetic field\cite{czajka2021oscillations}. These oscillations were reproduced by different groups\cite{bruin2022origin,lefranccois2023oscillations} but the origin is under hot debate. Czajka et al proposed that the observed oscillations as quantum oscillations of putative charge-neutral fermions akin to those produced by Landau quantization of electron states in a metal in the presence of magnetic fields\cite{czajka2021oscillations}. While others believed that the observed oscillatory features are the result of a sequence of magnetic field-induced magnetic phase transitions\cite{bruin2022origin,lefranccois2023oscillations,suetsugu2022evidence}. The experimental observation and understanding of the underlying physics are under debate for these two fascinating thermal transport properties, partially due to the materials issue of $\alpha$-RuCl$_3$\cite{lee2021}.

$\alpha$-RuCl$_3$ crystals are susceptible to stacking disorder due to the weak van der Waals bonding between the honeycomb layers. Stacking faults can form during crystal growth and sample handling. As demonstrated before\cite{cao2016low}, mechanical deformation can lead to magnetic anomalies in the temperature range 7\,K-14\,K. This property makes possible a comparative study of thermal transport properties of $\alpha$-RuCl$_3$ crystals with different amount/distribution of stacking disorder. In particular, this might lead to a control of the oscillatory features if they are indeed due to the field-induced magnetic phase transitions.

Motivated by this, we investigated the thermal transport properties of $\alpha$-RuCl$_3$ with different amount of stacking disorder introduced by mechanical deformation. Results from two crystals are presented in this paper. Crystal I shows a high Neel temperature near 13\,K. Sample II has less stacking disorder and shows two weak anomalies near 10\,K and 13\,K in addition to a dominant peak near 7.6\,K in the temperature dependence of specific heat. For both crystals, no obvious anomaly in the temperature and field dependence of thermal conductivity was observed to be associated with the magnetic order around 13\,K or 10\,K. Our results suggest that the oscillatory features of thermal conductivity should be innately tied to the zig-zag order phase at 7.6\,K. This observation is at odds with the idea that the magnetic transitions at, for example, 10\,K and 13\,K can contribute to the oscillatory features in $\alpha$-RuCl$_3$. Quite different planar thermal Hall effect was observed for those two crystals studied in this work. Overall, this work highlights the importance of controlling stacking disorder for more intrinsic thermal transport properties of $\alpha$-RuCl$_3$.

\begin{figure} \centering \includegraphics [width = 0.5\textwidth] {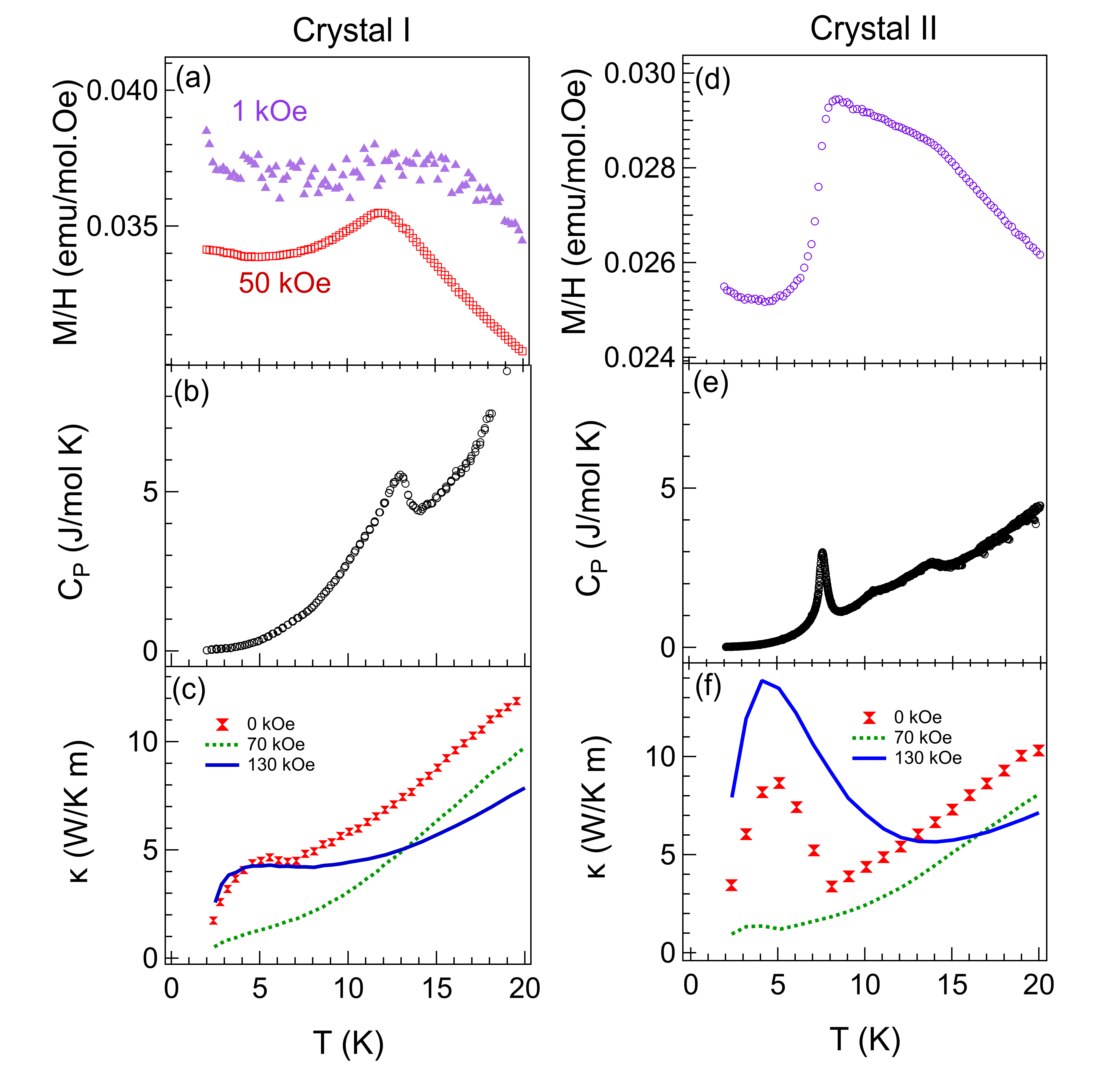}
\caption{(color online) Magnetization, specific heat, and thermal conductivity of (a-c) crystal I and (d-f) crystal II below 20\,K. (a) Temperature dependence of magnetization of crystal I measured with magnetic field applied along the zig-zag direction (perpendicular to the Ru-Ru bond). The magnetic data collected in a field of 1\,kOe is noisy because the crystal is only 0.15mg. We thus also show the data collected in a field of 50\,kOe that leads to a slightly lower T$_N$\cite{johnson2015monoclinic,bruin2022origin}. (b) Specific heat of crystal I showing a dominant lambda-type anomaly around 13\,K and a very weak anomaly barely observable near 7\,K. (c) Thermal conductivity of crystal I measured in different magnetic fields. (d) Temperature dependence of magnetization of crystal II with two anomalies at 7\,K and 14\,K. The data were collected in a magnetic field of 1\,kOe applied along the zig-zag direction. (e) Specific heat of crystal II showing a dominant lambda-type anomaly at 7.6\,K and two weak anomalies around 10\,K and 14\,K. Specific heat data for both samples were collected in zero magnetic field. (f) Thermal conductivity of crystal II measured in different magnetic fields. Both the heat current and magnetic field are along the zig-zag direction.}
\label{MagCpKappa-1}
\end{figure}

\section{Experimental details}
Millimeter sized $\alpha$-RuCl$_3$ crystals were grown using the conventional vapor transport technique with a temperature gradient of 250$\degree$C along the growth ampoule. About 0.3 gram of $\alpha$-RuCl$_3$ powder synthesized by reacting RuO$_2$ powder with AlCl$_3$-KCl salt\cite{yan2017flux} was sealed under vacuum inside of a fused quartz tube with an outer diameter of 16\,mm, a wall thickness of 1.0\,mm, and a length of 200\,mm.  The sealed ampoule was put inside of a two-zone tube furnace. The hot end with the starting powder was kept at 1000$\degree$C and the cold end at 750$\degree$C. After a week, the furnace was powered off to cool to room temperature. This kind of vapor transport growth results in plate-like crystals with in-plane dimension up to 4-5mm and thickness up to 0.2mm. Similar sized single crystals could also be obtained using self-selecting vapor transport technique\cite{yan2022self} when a cooling rate higher than 20$\degree$C/h is used.

Magnetic properties were measured with a Quantum Design (QD) Magnetic Property Measurement System in the temperature range 2.0\,K$\leq$T$\leq$\,30\,K. Specific heat data below 30\,K were collected using a QD Physical Property Measurement System (PPMS). Magnetic property and specific heat measurements confirm the as-grown crystals have only one single magnetic transition with the ordering temperature, T$_N$, of 7.6\,K. The well characterized crystals were then attached to kapton tape and the thickness was adjusted by peeling off part of the crystal with scotch tape. The kapton tape was then bent a couple of times to introduce stacking fault to the adhered $\alpha$-RuCl$_3$ crystals. This approach allows us to introduce stacking disorder without crumpling the crystals. As reported before\cite{cao2016low}, the stacking fault introduced this way will result in magnetic orders in the temperature range 7.6\,K - 14\,K. Intermediate magnetic measurements were performed during the bending process to monitor the anomalies in the temperature dependence of magnetization. In the end, the kapton tape was carefully removed.

Thermal transport measurements were carried out on a custom-built PPMS puck. We use Cernox as temperature sensors, and Model 336 Cryogenic Temperature Controller as thermometers. A one kilo-Ohm resistor was used as heater. We use gold wires (25 um) for thermal contact and manganin wires (25 um) for electric contact while minimizing thermal leakage. Contacts were made using silver paint from DuPont. All thermal transport measurements were carried out under high vacuum. The temperature dependence of magnetic susceptibility and specific heat below 20\,K was measured before and after measuring the thermal transport properties. This is to confirm that the thermal transport measurement doesn't introduce observable change to the magnetic properties and specific heat. It should be mentioned that magnetization, specific heat, and thermal transport properties for each type of crystals are all measured on exactly the same piece of crystal.

\section{Results and discussion}

Figure\,\ref{MagCpKappa-1} shows the temperature dependence of magnetization, specific heat, and thermal conductivity below 20\,K for two different crystals. From the magnetization and specific heat data shown in Fig.\,\ref{MagCpKappa-1}(a,b), crystal I shows a magnetic order at T$_N$\,=\,13\,K. A weak feature barely observable near 7\,K in (b) indicates the presence of a small fraction of original nondeformed phase. Crystal II has less amount of stacking fault and three anomalies can be observed in specific heat data shown in Fig.\,\ref{MagCpKappa-1}(e). The dominant one is found at 7.6\,K. These two crystals enable us to investigate how the stacking disorder affects the thermal transport properties. We tried to obtain a crystal with only one magnetic order with T$_N$ around 10\,K but failed.

The response to the magnetic order of longitudinal thermal conductivity of $\alpha$-RuCl$_3$ has been reported by many groups\cite{hentrich2018unusual, hirobe2017magnetic,leahy2017anomalous,kasahara2022quantized,lefranccois2022evidence}. Despite the variation of the magnitude, thermal conductivity data reported by different groups show similar temperature dependence: thermal conductivity is enhanced upon cooling through T$_N$ and shows a peak around 5\,K. This kind of recovery of lattice thermal conductivity upon cooling through a magnetic order has been observed in many other systems with strong spin-lattice coupling. From a simple analogy, one would expect thermal conductivity to resurge upon cooling below 13\,K for crystal I and show some weak anomalies near 10\,K and 13\,K for crystal II. Figures\,\ref{MagCpKappa-1} c and f show the temperature dependence of thermal conductivity of both crystals. Surprisingly, no obvious anomaly was observed above 7.6\,K in zero magnetic field. Both crystals show a recovering of thermal conductivity when cooling below 7.6\,K. This feature is much more dramatic for crystal II than I. Around 5\,K where the thermal conductivity peaks, thermal conductivity of crystal II is about twice of that for crystal I. This is consistent with the fact that crystal II has a much stronger response in magnetization and specific heat near T$_N$ at 7.6\,K. The absence of observable anomaly around 13\,K and the recovery upon cooling through 7.6\,K for both crystals indicate that the magnetic order at 7.6\,K has a more dramatic effect on the longitudal thermal conductivity of $\alpha$-RuCl$_3$. Despite a bulk behavior of the magnetic order at 13\,K for crystal I determined from the temperature dependence of magnetization and specific heat, this magnetic order shows little effect on the longitudal thermal conductivity. This interesting observation inspires one to study how the thermal conductivity responds to a high in-plane magnetic field.

Figures\,\ref{MagCpKappa-1}c and f also show the temperature dependent longitudal thermal conductivity measured in magnetic fields of 70\,kOe and 130\,kOe applied along the zig-zag direction (perpendicular to the Ru-Ru bond). In the spin polarized state, crystal II shows a much higher thermal conductivity at low temperatures. This is consistent with the larger increase of thermal conductivity below T$_N$ for crystal II when measured in zero magnetic field. In an applied magnetic field of 70\,kOe, the long range magnetic order is suppressed and so does the lattice heat transport.

\begin{figure} \centering \includegraphics [width = 0.5\textwidth] {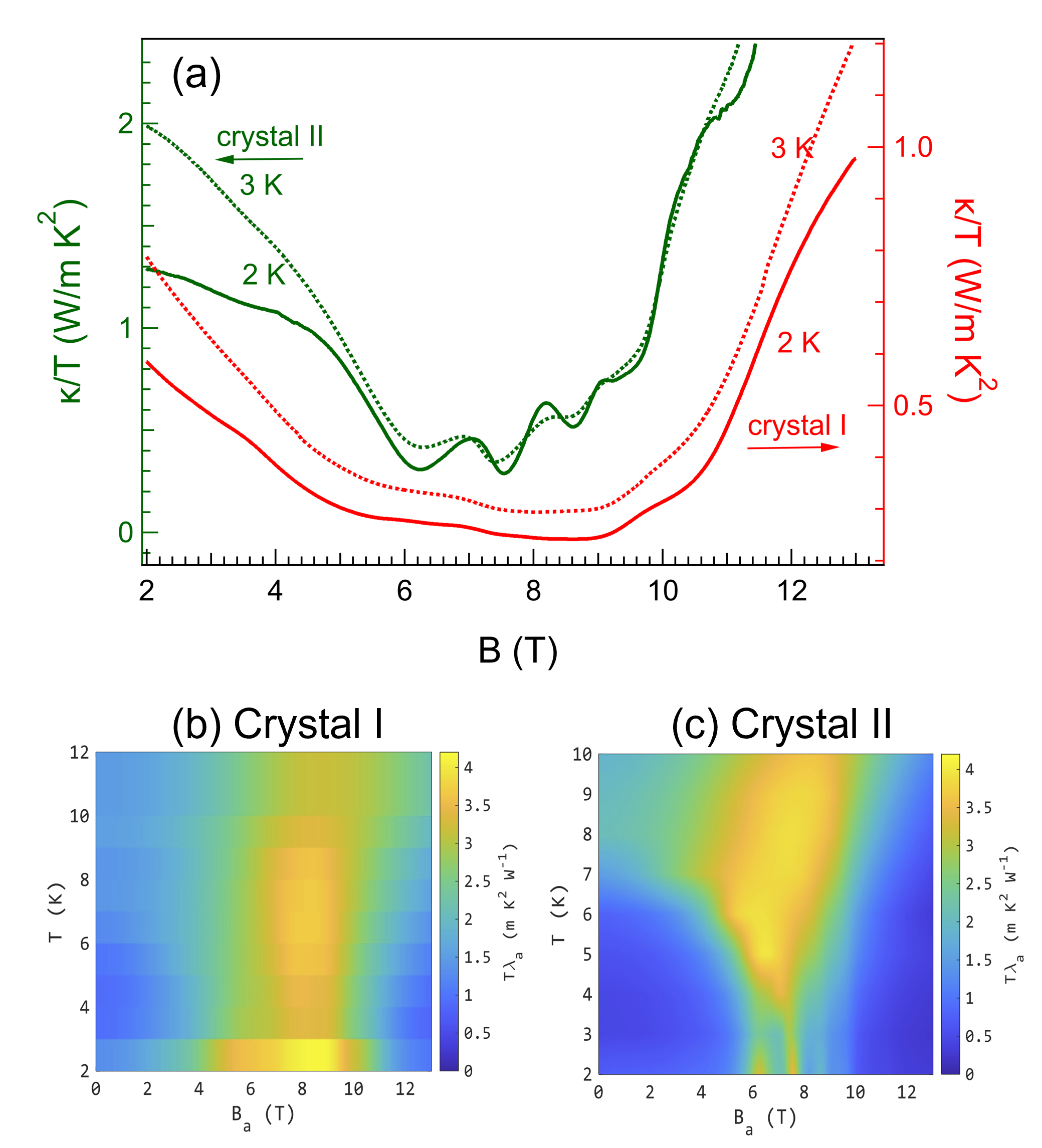}
\caption{(color online) Field dependence of longitudal thermal conductivity $\kappa_{xx}$/T. (a) $\kappa_{xx}$/T at low temperatures. (b, c) Color plot of $\kappa_{xx}$/T in the temperature-field space studied in this work. Both the heat current and magnetic field are along the zig-zag direction (or perpendicular to the Ru-Ru bond).}
\label{kxx-1}
\end{figure}

Figure\,\ref{kxx-1} (a) shows the field dependence of thermal conductivity at low temperatures for both crystals. For crystal II, the oscillatory features can be well resolved. At 2\,K, the minima in magnetothermal conductivity show up at 62, 75, 86, 96, and 112\,kOe. The presence of the oscillatory features and the critical fields agree well with those observed in our $\alpha$-RuCl$_3$ crystals with minimal amount of stacking disorder (reported separately) and also those reported previously by other groups\cite{bruin2022origin}. We measured the field dependence of thermal conductivity of some other pieces of crystals like crystal II but with different amount/distribution of stacking disorder. All these crystals show a dominant magnetic transition around 7\,K and they are different by showing weak anomalies of different magnitudes around 10\,K and/or 14\,K in specific heat curves. All these crystals show the minima in magnetothermal conductivity at the same critical fields. We don't see any correlation between these critical fields and the magnitude of specific heat anomalies around 10\,K and 14\,K. The stacking disorder doesn't seem to affect these critical fields but does affect the magnitude of thermal conductivity.

\begin{figure*} \centering \includegraphics [width = 0.9\textwidth] {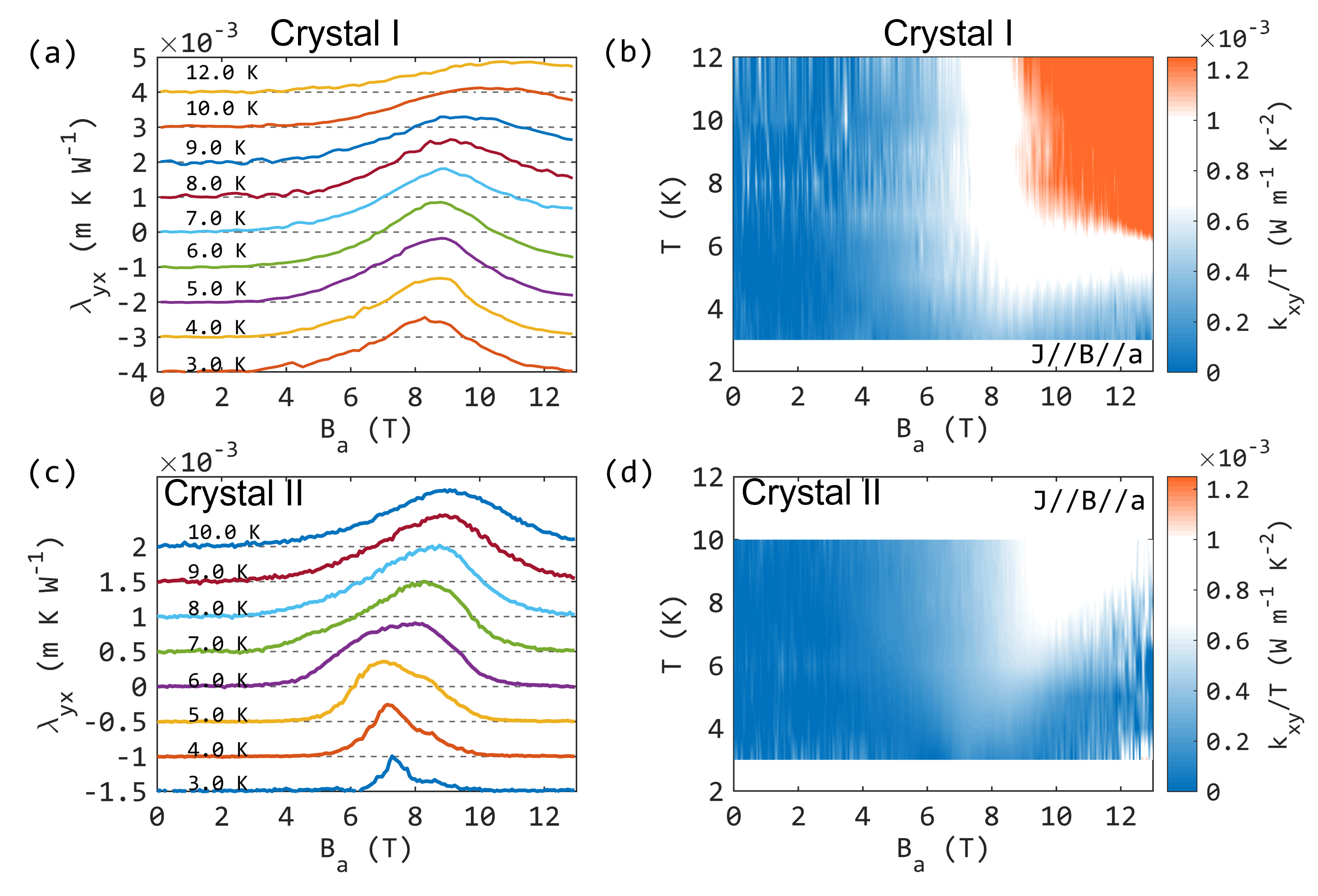}
\caption{(color online) Planar thermal Hall response with magnetic field and heat current parallel to the zig-zag direction (or perpendicular to the Ru-Ru bond). (a, c) Field dependence of thermal Hall resistivity at different temperatures for crystal I (a) and II (c). The horizontal dashed lines show the zero base line for each temperature. (b,d) Color plot of thermal Hall conductivity $\kappa_{xy}$/T for crystal I (b) and II (d). The white region shows the $\kappa_{xy}$/T within $\pm$20\% of $\kappa_{HQ}$/T.}
\label{kxy-1}
\end{figure*}

In contrast to the pronounced oscillating features for crystal II, only rather weak features are observed for crystal I in the magnetothermal conductivity curves in the field range of 40\,kOe-120\,kOe. These weak features follow the similar field dependence as for those oscillatory features for crystal II. This similar field dependence suggests for crystal I that (1) those weak features may come from the residual small fraction of the original phase with T$_N$=7.6\,K, and (2) the magnetothermal conductivity is also dominated by the T$_N$=7.6\,K phase and there is no unique features that could be attributed to other magnetic phases with a T$_N>$7.6\,K. This is quite different from our expectation. According to previous studies of the magnetic order in applied magnetic field\cite{johnson2015monoclinic,bruin2022origin}, the T$_N$=13\,K magnetic order should be suppressed by an in-plane magnetic field near 90\,kOe. Around this magnetic field, one would expect a well defined feature in the magnetothermal conductivity curve of crystal I if the oscillatory features observed by different groups come from the field-induced phase transitions. Unfortunately, we didn't observe any feature dominated by the magnetic phase with T$_N$=13\,K in our magnetothermal conductivity data for crystal I. Since the stacking disorder is introduced by bending the crystals after growth, one might wonder whether thermal conductivity responds to the stacking induced magnetic orders with T$_N>$7.6\,K in a wider temperature and/or field range. If this is true, one doesn't expect to see well defined features from field-induced phase transitions in the magnetothermal conductivity curves for crystal II.

Figures\,\ref{kxx-1}b and c show the color plot of thermal resistivity T$\lambda$ over the whole temperature-field  space investigated in this work. The oscillatory features cannot be well resolved any more above 4\,K for crystal II and the overall feature shown in Fig.\,\ref{kxx-1}c agrees well with what's reported by Czajka et al\cite{czajka2022planar}. For crystal I with the dominant magnetic order at 13\,K, the weak anomalies observed at 2\,K also disappear above 4\,K and the field range, in which T$\lambda$ shows a maximum, exhibits little change with increasing temperature.

The above results suggest that the magnetic phases with T$_N$=10\,K and 13\,K don't produce observable signatures in thermal conductivity. However, our planar thermal Hall data suggest that they or the stacking disorder can have a significant effect on thermal Hall effect of $\alpha$-RuCl$_3$. Figures\,\ref{kxy-1} (a, c) show the field dependence of thermal Hall resistivity measured at different temperatures up to 12\,K. For crystal I, a nonzero thermal Hall resistivity was observed in a wide field range 40\,kOe-120\,kOe and a broad peak centering around 850\,kOe was observed. This magnetic field is similar to that required to suppress the magnetic order at T$_N$=14\,K\cite{johnson2015monoclinic,bruin2022origin}. With increasing temperature, this broad peak moves toward higher magnetic field. For crystal II, in addition to an anomaly around 85\,kOe, the dominant feature centers around 72\,kOe.  The evolution with temperature of this dominant feature follows the change of T$_N$ in applied magnetic field and cannot be well resoved above 8\,K. The feature around 85\,kOe shows the same temperature dependence as the main feature for crystal I shown in (a). Figures\,\ref{kxy-1} b and d show the color plot of thermal Hall $\kappa_{xy}$/T. The white region shows the $\kappa_{xy}$/T within $\pm$20\% of the half quantized thermal Hall conductivity $\kappa_{HQ}$/T. The L-shaped white region in Fig.\,\ref{kxy-1}(a) for crystal I resembles that previously reported by Bruin et al \cite{bruin2022robustness}. The white region runs vertically from 12\,K to about 6\,K at around 80\,kOe and then continues horizontally from around 70\,kOe to at least 130\,kOe at about 6\,K. For crystal II, $\kappa_{xy}$/T reaches to $\kappa_{HQ}$/T only in the high temperature high field regime. This behavior is more in line with what's observed by Czajka et al\cite{czajka2022planar}.

\section{Summary}

In summary, we report the thermal transport properties of two $\alpha$-RuCl$_3$ with different amount of stacking disorder introduced by mechanical deformation. Crystal I shows only one magnetic order around 13\,K. Crystal II has smaller amount of stacking disorder and specific heat data show a dominant transition at 7.6\,K and two weak anomalies around 10 and 13\,K. No obvious anomaly in the temperature and field dependence of longitudal thermal conductivity was observed to be associated with the magnetic order around 10\,K or 13\,K. Similar oscillatory features in the field dependence of thermal conductivity were observed in all crystals that show a dominant magnetic order around 7\,K. For crystal I, an L-type shape was observed for the region in temperature-field space in which thermal Hall conductivity $\kappa_{xy}$/T is within $\pm$20\% of the half quantized thermal Hall conductivity $\kappa_{HQ}$/T. $\kappa_{xy}$/T for crystal II reaches $\kappa_{HQ}$/T only in the high field and high temperature regime. Our observation suggests that the oscillatory feature may be an unique character of the magnetic phase with T$_N$ near 7\,K instead of resulting from field-induced magnetic phase transitions. Our results also show that the planar thermal Hall effect depends on the stacking disorder which deserves further careful investigation.

\section{Acknowledgment}
HZ, SN, MM, and JY were supported by the U.S. Department of Energy, Office of Science, National Quantum Information Science Research Centers, Quantum Science Center. AM, HM, BS, and DM were supported by the US Department of Energy, Office of Science, Basic Energy Sciences, Materials Sciences and Engineering Division.

 This manuscript has been authored by UT-Battelle, LLC, under Contract No.
DE-AC0500OR22725 with the U.S. Department of Energy. The United States
Government retains and the publisher, by accepting the article for publication,
acknowledges that the United States Government retains a non-exclusive, paid-up,
irrevocable, world-wide license to publish or reproduce the published form of this
manuscript, or allow others to do so, for the United States Government purposes.
The Department of Energy will provide public access to these results of federally
sponsored research in accordance with the DOE Public Access Plan (http://energy.gov/
downloads/doe-public-access-plan).

\section{references}
%

\end{document}